\documentclass[numbers,showpacs,nopreprint,superscriptaddress]{jasatex}

\usepackage{amsmath,amscd,amssymb}
\usepackage{bm}
\usepackage{graphicx}
\usepackage[tight]{subfigure}
\usepackage{units}
\usepackage{array,booktabs}
\usepackage{amssymb}
\usepackage{color}

\begin{document}

% ================================================================================================

\title{Acoustic radiation force and torque on an absorbing compressible particle in an inviscid fluid}
\author{Glauber T. Silva}
\email{glauber@pq.cnpq.br}
\affiliation{Physical Acoustics Group, Instituto de F\'isica, 
	Universidade Federal de Alagoas, Macei\'o, AL 57072-900, Brazil.}

% ================================================================================================
\begin{abstract}
Exact formulas of the acoustic radiation force and torque exerted by an arbitrary time-harmonic wave 
on an absorbing compressible particle that is suspended in an inviscid fluid are presented.
It is considered that the particle diameter is much smaller than the incident wavelength, i.e. the so-called Rayleigh
scattering limit.
Moreover, the particle absorption assumed here is due to the attenuation of compressional waves only.
Shear waves inside and outside the particle are neglected, since 
the inner and outer viscous boundary layer of the particle are supposed to be much smaller than the particle radius.
The obtained radiation force formulas are used to establish the trapping conditions of a particle by a single-beam
acoustical tweezer based on a spherically focused ultrasound beam.
In this case, it is shown that the particle absorption has a pivotal role in single-beam trapping at the transducer focal region.
Furthermore, it is found that only the first-order Bessel vortex beam can generate the radiation torque on a small particle.
In addition,  numerical evaluation of the radiation force and torque exerted on a benzene and an olive oil droplet 
suspended in water are presented and discussed.
\end{abstract}
 % replace with appropriate codes (Section 43 Acoustics):  
\maketitle

%\pacs{} % replace with appropriate codes (Section 43 Acoustics):  

 \section{Introduction}

An increasing interest on acoustic radiation force has arisen after
the concept of acoustical tweezers was developed.
An acoustical tweezer can be accomplished by means of an ultrasound
standing wave~\cite{wu:2140} or a focused beam~\cite{lee:073701}. 
Furthermore, acoustophoretic devices are also relies on the acoustic 
radiation force actuating on microsized particles~\cite{evander:4667,ding:11105,foresti:12549}.
A key point in developing and enhancing the use of acoustical tweezers and acoustophoretic devices is to
understand how an employed acoustic wave generates radiation force and torque on a small particle 
(i.e. a particle whose radius is much smaller than the incident wavelength).
\medskip

The first study on the radiation force exerted on spherical particle  was presented 
in a seminal work by King~\cite{king:212}.
An extension of this work considering an incident spherical wave  was provided
by Emblenton~\cite{embleton40}. 
It was noticed that the particle can be either attracted to or repulsed by the wave source depending on
their relative distance.
Effects of particle compressibility on the radiation force was accounted 
by Yosioka~\emph{et al}.~\cite{yosioka:167}
Subsequently, Gorkov~\cite{gorkov:773} derived, based on a fluid dynamics approach, a general
radiation force formula exerted on a particle by a plane wave and any stationary acoustic wave.
Nyborg~\cite{nyborg:947} has also considered the radiation force exerted by a spherical wave on a rigid sphere.
He showed that the traveling part of the spherical wave can be described with Gorkov's theory by
adding a correction term.
Wu~\emph{et al.}~\cite{wu:997} analyzed the radiation force produced by  a  Gaussian beam and  
a focused piston  on an on-axis particle.
Recently, Marston~\cite{marston:3518} obtained the radiation force generated by the interaction of a zero-order Bessel beam 
and an on-axis particle. 
He gave the conditions to produce a negative force (i.e. opposite to the beam's propagation direction) 
on the particle. 
The radiation force of a Bessel beam exerted on a small particle has been also investigated in Ref.~~\onlinecite{baresch2013a}.
Moreover,
Aazarpeyvand~\cite{azarpeyvand:4337} has performed a theoretical study on the radiation force exerted on an on-axis porous aluminum sphere.
In this case, the increase of porosity degrades the radiation force on a small particle.

Commonly, acoustic particle trapping is described through Gorkov's theory~\cite{gorkov:773}.
In turn, this theory is valid when the Stokes' boundary layer around the particle satisfies 
$\delta_0 = \sqrt{ 2 \nu_0 /\omega}\ll a$,
where $a$ is the particle radius, $\nu_0$ is the kinematic viscosity of the host fluid 
and $\omega$ is the angular frequency of the wave.
When $\delta_0 \sim a$, effects of the host fluid viscosity become relevant and the radiation force may considerably deviate from 
 from inviscid theories~\cite{doinikov:1,settnes:016327}.
It should be further noticed that if the particle is made of a viscous fluid,
a shear wave may develop inside it.
The propagation of this  wave can be confined within an internal viscous boundary layer, which is given by~\cite{epstein:553} $\delta_1 = \sqrt{ 2 \nu_1 /\omega}$, where $\nu_1$ is the kinematic viscosity of the particle.
Here, we will consider particles for which $\delta_1\ll a$.
Thus, shear wave propagation effects on the radiation force and torque will be neglected.

Another important phenomenon in the acoustic radiation force is the  absorption of longitudinal waves inside the  particle.
%All previous theories reported here did not consider this effect.
Radiation force on an absorbing  particle was first discussed by Westervelt~\cite{westervelt:312}.
He established a relation between 
the radiation force on a sphere due to  a plane wave with the scattering and the absorption cross-sections. 
A more extensive analysis of this problem was performed by 
L\"ofstedt~\emph{et al.}~\cite{lofstedt:2027}
in which it is shown that absorption enhances the radiation force. 
So far,  the radiation force theory in the Rayleigh scattering limit involve  a symmetry consideration
between the particle and the incident wave, i.e. the on-axis configuration.
Furthermore, Zhang~\emph{et al.}~\cite{zhang:035601} developed 
a geometric interpretation of the radiation force
exerted by a zero-order Bessel beam on an on-axis sphere in terms of the absorption, 
scattering and extinct cross-sections.
A similar study on a broader class of acoustic beams was also performed by this 
group~\cite{zhang:EL329,zhang:1610}.
The interaction of acoustic wave with an absorbing particle may induce a radiation torque.
Hefner~\emph{et al.}~\cite{hefner:3313} showed that an axial acoustic radiation torque 
can be developed on an absorbing target by a paraxial vortex beam.
Moreover, the axial radiation torque on an axisymmetric object suspended in an inviscid fluid was theoretically studied 
by Zhang~\emph{et al.}~\cite{zhang:065601}
A general model for the three-dimensional radiation torque exerted by an acoustic beam of arbitrary wavefront 
was provided by Silva~\emph{et al.}~\cite{silva:54003}

The purpose of this article is to provide exact formulas of the acoustic radiation force and torque 
generated  by any time-harmonic beam acting on an absorbing particle placed anywhere in an inviscid fluid.
The radiation force formula is derived
stemming from the partial-wave expansion method
up to the quadrupole moment of the incident wave~\cite{silva:3541, silva:54003}.
Whereas the expression for the radiation torque is derived using the theory developed in Ref.~~\onlinecite{silva:54003}.
The ultrasound absorption is assumed to obey a frequency power-law model~\cite{szabo:491}.
Nevertheless, shear wave propagation inside and outside the particle is not taken into account, 
since  $\delta_0, \delta_1\ll a$  is considered.
The developed theory is applied to the analysis of the radiation force due to a spherically focused beam on a 
benzene and an olive oil droplet.
Results show that transverse trapping depends on the particle's compressibility and density with respect to
the host fluid, while axial trapping also reckons on the particle absorption. 
In addition, we derive the radiation force and torque produced by a Bessel vortex beam 
in the on-axis configuration.
It is shown that only the first-order Bessel vortex beam can generate radiation torque 
in the Rayleigh scattering limit.

\section{Model equations}

Consider  an acoustic wave 
propagating in a inviscid fluid of ambient density $\rho_0$ and speed of sound $c_0$.
The wave propagation is characterized by the excess of pressure $p$,
the density $\rho$, and the the fluid element velocity ${\bm v}$.
These acoustic fields are functions of  the position vector ${\bf r}$ and the time $t$.
For low-amplitude  waves, i.e. $|p|^2/\rho_0 c_0^2\ll 1$, the acoustic fields  
satisfy the fluid dynamics equations in linear approximation,~\cite{pierce:1989}
\begin{align}
\label{ceq1}
\frac{\partial \rho}{\partial t} + \rho_0 \nabla \cdot {\bm v} &= 0,\\
\label{ceq2}
\rho_0\frac{\partial {\bm v}}{\partial t} + \nabla p &= 0,\\
\label{ceq3}
p &= c_0^2 \rho.
\end{align}
By taking the divergence of Eq.~(\ref{ceq2}) and substituting the result in Eq.~(\ref{ceq1}),
we obtain, after eliminating the density, the wave equation for the pressure as follows
\begin{equation}
\label{wave_eq}
 \left(\nabla^2- \frac{1}{c_0^2}\frac{\partial^2}{\partial t^2}\right) p = 0.
\end{equation}
In the foregoing discussion of radiation force and torque, we will consider only time-harmonic acoustic waves, whose
pressure are of the form $p({\bm r}) e^{-i \omega t}$.
Hence, considering a time-harmonic pressure in Eq.~(\ref{wave_eq}) we obtain the Helmholtz equation,
\begin{equation}
 (\nabla^2+k^2) p({\bm r}) = 0,
\end{equation}
where $k=\omega/c_0$ is the wavenumber.
The term $e^{-i \omega t}$ was suppressed for simplicity.

\section{Rayleigh scattering}

Assume that an incident wave  is scattered by a  compressional fluid particle of  radius $a$, density $\rho_1$, 
and speed of sound $c_1$.
The particle is centered in the origin of the coordinate system.
The total pressure amplitude  in the host fluid is described by $p_\text{i}({\bm r})+p_\text{s}({\bm r}) $,
where $p_\textrm{i}$ and $p_\textrm{s}$ represent the incident and the scattered pressure, respectively.
The scattered pressure
is expanded in a partial-wave series in spherical coordinates $(r,\theta,\varphi)$ as follows~\cite{silva:298}
\begin{equation}
p_\text{s} = p_0 \sum_{n=0}^\infty\sum_{m=-n}^m s_n a_n^m h_n^{(1)}(k r)Y_n^m(\theta,\varphi),
\label{ps}
\end{equation}
where $p_0$ is  the peak pressure magnitude of the incident wave, $k=\omega/c_0$ is the  wavenumber,
$h_n^{(1)}$ the $n$th-order spherical Hankel function of first-type,
and $Y_n^m$  is the $n$th-order and $m$th-degree
spherical harmonic.
The quantities $a_n^m$ and $s_n$ are the beam-shape and the scaled scattering coefficients, respectively.
Note that Eq.~(\ref{ps}) satisfies the Sommerfeld radiation condition. 
The beam-shape coefficients are the weights of the partial-waves in the incident beam expansion~\cite{silva:298}.
They will be determined in terms of the incident pressure and fluid element velocity 
up to the quadrupole approximation in Appendix A.
%Note that $a_n^m = 0$ if $n<0$ or $|m|>n$.

The scaled scattering coefficient is obtained by applying the continuity condition of pressure and particle 
velocity  across the particle surface at $r=a$.
These conditions lead to 
\begin{equation}
\label{sn}
s_n = -
\det
\left[
\begin{matrix}
 \gamma j_n(k a) & j_n(\kappa_1 a)\\
j_n'(k a) & j_n'(\kappa_1 a)
\end{matrix}
\right]  \det
\left[
\begin{matrix}
 \gamma h_n^{(1)}(k a) & j_n(\kappa_1 a)\\
{h_n^{(1)}}'(k a) & j_n'(\kappa_1 a)
\end{matrix}
\right]^{-1},
\end{equation}
where
$\kappa_1$ in the inner wavenumber of the particle, $\gamma= \rho_0 \kappa_1  / (\rho_1 k) $, and
the prime symbol indicates differentiation.
In the Rayleigh scattering regime, the particle is much smaller than the incident wavelength
or $ka\ll 1$.
In this regime, the monopole and the dipole scattering coefficients are $O[(ka)^6+i(ka)^3]$, while
$s_n = O[(ka)^{4n+2}+i(ka)^{2n+1}]$ for $n>1$. 
Thus, $ka= 0.3$ can be regarded as an upper-limit for the Rayleigh scattering approach, 
because it renders an error above
$9\%$ on both the real and the imaginary parts of the scaled scattering coefficient $s_n$.

It is assumed that the ultrasound absorption within the particle obeys a frequency power-law as follows~\cite{szabo:491}
$\alpha_\upsilon = \alpha (\omega/2 \pi)^\upsilon,$
where $\alpha$ is the absorption coefficient and $0<\upsilon\le 2$. 
Thus, the inner wavenumber is expressed as
\begin{equation}
\label{k1}
\kappa_1  =k_1 + i \alpha_\upsilon,
\end{equation}
where $k_1=\omega/c_1$.
We may neglect shear wave propagation effects inside the particle, when the inner viscous boundary layer is much smaller 
the particle radius, $\delta_1/a \ll 1$.
Thus, we assume as an upper-limit $\delta_1=0.1 a$.
Since we have established that $ka\le 0.3$ and given that
$\delta_1=\sqrt{2 \nu_1/\omega}$, then
the particle size factor $ka$ should satisfy
\begin{equation}
\frac{10 \sqrt{2 \nu_1 \omega}}{c_0} \le ka \le 0.3.
\label{ineq_a}
\end{equation}

We now expand the scaled scattering coefficients $s_0$ and $s_1$ 
in Eq.~(\ref{sn}) for $a\rightarrow 0$ and $ka\ll \rho_1/\rho_0$.
It is useful to define the dimensionless absorption coefficient $\tilde{\alpha}_\upsilon=\alpha_\upsilon/k_1$.
Our analysis is limited to  weak-absorption, i.e. $ \tilde{\alpha}_\upsilon \ll 1$ is assumed.
Using \textsc{Mathematica} software~\cite{mathematica8}, we find the first relevant terms of monopole and
dipole scattering coefficients as
\begin{align}
\label{s0}
 s_0 &= 2(f_0-1)\tilde{\alpha}_\upsilon \frac{(ka)^3}{3}- f_0^2\frac{(ka)^6}{9} - i f_0 \frac{(ka)^3}{3},\\
 %2(f_0-1)\biggl[ \left(\frac{c_0}{c_1}\right)^2 \\
%&-10(f_0 - 1) - 15\biggr]
%s_0 &= -\frac{2  \rho_0 c_0^ 2 \alpha_\upsilon\omega^{\upsilon - 1}}{ \rho_1 c_1 } \frac{(ka)^3}{3}
 %-  f_0^2\frac{(ka)^6}{9} 
s_1 &=  2\frac{(f_0-1)\tilde{\alpha}_\upsilon}{( \tilde{\rho}^{-1}_1 + 2)^2}  \frac{(ka)^5}{5}
  - f_1^2 \frac{(ka)^6}{36} +i f_1\frac{(ka)^3}{6},
  \label{s1}
\end{align}
where $\tilde{\rho}_1=\rho_1/\rho_0$, $f_0 = 1 - \rho_0 c_0^2/\rho_1 c_1^2$ and $
 f_1 = 2(\rho_1 - \rho_0)/(2 \rho_1 + \rho_0)$ are the monopole and the dipole scattering factors, respectively.
 Inasmuch as we are considering weak-absorption, i.e. $\tilde{\alpha}_\upsilon\ll 1$,
 we have neglected terms involving $\tilde{\alpha}_\upsilon^m$, $m=2,3,\dots$ in Eqs.~(\ref{s0}) and (\ref{s1}), 
 %Equations~(\ref{s0}) and (\ref{s1}) will enable us to decompose the radiation force as a sum of two components, 
 %one related to the particle absorption and the other as though the particle was lossless.
 
\section{Acoustic radiation force}

The linear momentum carried by the incident wave is transferred
to the suspended particle generating the so-called acoustic radiation force.
Based on the partial-wave expansion of the incident and the scattered fields,
it has been shown that the radiation force exerted on a sphere by a time-harmonic beam with arbitrary wavefront
is given by~\cite{silva:3541}
\begin{equation}
{\bm F}= \frac{\pi a^2 I_0}{c_0}{\bm Y},
\end{equation}
where $I_0 = p_0^2/2\rho_0 c_0$ is the averaged incident intensity and
${\bm Y} $ is the dimensionless radiation force vector, with
$Y_x$, $Y_y$, and $Y_z$ being its Cartesian components.
In turn, this vector is expressed in terms of the beam-shape $a_n^m$ and the scaled scattering coefficient $s_n$~\cite{silva:1207}.
For the Rayleigh scattering limit, only $s_0$ and $s_1$ are relevant as previously discussed.
Therefore, keeping only these terms in [Eqs.~(1)-(3),~~\cite{silva:1207}], one finds
that the Cartesian components of the dimensionless radiation force are given by
\begin{align}
\nonumber
&\mbox{} Y_x + iY_y =\frac{i}{2\pi(ka)^2}\biggl[ 
\sqrt{\frac{2}{3}} \bigl[ (s_0 + s_1^* + 2 s_0 s_1^*)a_0^0 a_{1}^{1*}  \\
\nonumber
&+ (s_0^* + s_1 + 2 s_0^* s_1 ) a_0^{0*} a_{1}^{-1}  \bigr]
+ \sum_{m=-1}^1 \sqrt{\frac{(2+m)(3+m)}{15}} \\
\label{Yxy2}
& \times
 \bigl( s_1 a_1^m a_{2}^{m+1*}   + s_1^*  a_1^{-m*} a_{2}^{-m-1}  \bigr) \biggr],\\
 \nonumber
&\mbox{} Y_z= \frac{1}{\pi(ka)^2}\textrm{Im} \biggl[\sqrt{\frac{1}{3}} (s_0 + s_1^* + 2 s_0 s_1^*) a_0^0 a_1^{0*}\\
&+ \sum_{m=-1}^1 \sqrt{\frac{(2-m)(2+m)}{15} } s_1 a_1^m a_{2}^{m*} \biggr].
\label{Yz2}
\end{align}
where  the symbol $^*$ means complex conjugation and `Im' signifies the imaginary-part of.
We stress here that $Y_x$, $Y_y$, and $Y_z$ are real quantities.
Note that the quadrupole moment of the incident beam $a_2^m$ is necessary to compute the radiation force.
This result has been also noticed in the axial radiation force exerted on a particle by a Bessel beam~\cite{marston:3518}.

To obtain the radiation force, we substitute the equations in~(\ref{a22}) into 
Eqs.~(\ref{Yxy2}) and (\ref{Yz2}).
Furthermore, we consider the relation 
\begin{equation}
\label{nabla_v}
\nabla\cdot {\bm v}_\text{i} = \frac{i k}{\rho_0 c_0} p_\text{i},
\end{equation}
which is derived by combining of Eq.~(\ref{ceq3}) and Eq.~(\ref{ceq1}).
Hence, we find the radiation force as
\begin{align}
\nonumber
{\bm F} &= - \frac{2 \pi }{k^2 c_0} \text{Re} \bigg[\frac{3is_1}{k} \rho_0 c_0  
{\bm v}_\text{i}({\bm 0})\cdot\nabla {\bm v}_\text{i}\mbox{}^*({\bm 0}) \\
&+ (s_0 + 2s_0 s_1^*)p_\text{i}({\bm 0}) {\bm v}_\text{i}\mbox{}^*({\bm 0})\bigg],
\label{rf_rayleigh1}
\end{align}
where  `Re' means the real-part. In Cartesian coordinates, $\nabla {\bm v}_{\text{i}}=(\nabla  v_{\text{i},x},
\nabla  v_{\text{i},y},\nabla v_{\text{i},z})$. 
It should be remarked that in Eq.~(\ref{rf_rayleigh1}) the acoustic fields are evaluated in the particle center, 
 ${\bm r} = {\bm 0}$.

 To further develop Eq.~(\ref{rf_rayleigh1}) consider the identity
\begin{equation}
\nabla \cdot {\bm v}_\text{i} {\bm v}_\text{i}\mbox{}^* 
= {\bm v}_\text{i}\cdot \nabla{\bm v}_\text{i}\mbox{}^*+ {\bm v}_\text{i}\mbox{}^* (\nabla \cdot {\bm v}_\text{i}).
\label{nabla_vv}
\end{equation}
Considering Eq.~(\ref{nabla_v}) the last term in the right-hand side of Eq.~(\ref{nabla_vv}) becomes 
$i k p_\text{i}{\bm v}_\text{i}^*/c_0 $.
Thus, the acoustic radiation force in Eq.~(\ref{rf_rayleigh1}) turns to
\begin{align}
\nonumber
{\bm F} &= - \frac{2 \pi  }{k^2 c_0} \text{Re} \bigg[\frac{3i\rho_0 c_0 s_1}{k} 
\nabla \cdot {\bm v}_\text{i} {\bm v}_\text{i}\mbox{}^*({\bm 0})
+ (s_0+ 3s_1 + 2s_0 s_1^*)\\
&\times p_\text{i}({\bm 0}) {\bm v}_\text{i}\mbox{}^*({\bm 0})\bigg].
\label{rf_rayleigh2}
\end{align}
Using the relation~\cite{westervelt:312}
\begin{equation}
\textrm{Re}\left[ \nabla \cdot\rho_0 {\bm v}_\text{i} {\bm v }_\text{i}\mbox{}^* \right]= 
\nabla \left( \frac{\rho_0  |{\bm v}_\text{i}|^2} {2} - \frac{ |p_\text{i}|^2}{2\rho_0 c_0^2}\right),
\end{equation}
along with Eqs.~(\ref{s0}) and (\ref{s1})
into Eq.~(\ref{rf_rayleigh2}), we obtain
\begin{align}
\nonumber
{\bm F} &= -\nabla U({\bm 0})
+\biggl[\frac{4\pi a^2  }{9 c_0}\left(f_0^2 + f_0f_1 + \frac{3f_1^2}{4} \right) 
 (ka)^4 \\
 \nonumber
&+ \frac{8\pi k a^3 (1-f_0) \tilde{\alpha}_\upsilon}{3 c_0}\biggr]
  \overline{\bm I}({\bm 0}) + \biggl[
  \frac{12\pi(f_0-1)\tilde{\alpha}_\upsilon}{5( \tilde{\rho}^{-1}_1 + 2)^2} k^2 a^5 \\
  &-  \frac{\pi  f_1^2 k^3a^6}{6}\biggr]
  \textrm{Im}[ \nabla \cdot \rho_0 {\bm v}_\text{i} {\bm v }_\text{i}\mbox{}^*({\bm 0})],
% &+\frac{8 \pi a^3 \rho_0 \alpha_\upsilon \omega^{\upsilon} }{3 \rho_1c_1}\biggr]
%   \overline{\bm I}({\bm 0}) -  \frac{\pi  f_1^2 k^3a^6}{6}\textrm{Im}[ \nabla \cdot \rho_0 {\bm v}_\text{i} {\bm v }_\text{i}^*({\bm 0})],
\label{rf_rayleigh3}
\end{align}
where
\begin{equation}
\label{potential_function}
 U =   \pi a^3  \left(f_0 \frac{|p_\text{i}|^2}{3 \rho_0 c_0^2} - f_1 \frac{\rho_0 |{\bm v}_\text{i}|^2}{2}\right)
\end{equation}
is the radiation force potential function~\cite{gorkov:773}
and $\overline{\bm I}= (1/2)\text{Re}\{p_\text{i} {\bm v}_\text{i}^*\}$ is the incident  intensity averaged in time.
Note also that Eq.~(\ref{rf_rayleigh3}) has been obtained in a different format in Ref.~~\onlinecite{sapozhnikov:661}.
According to Eq.~(\ref{rf_rayleigh3}), 
the radiation force can be decomposed into three contributions, namely gradient, scattering,
and absorption components.

The gradient radiation force is given by
\begin{equation}
\label{grad}
 {\bm F}_\text{grad} = - \nabla U({\bm 0}),
\end{equation}
This force was first obtained by Gorkov.~\cite{gorkov:773}
The gradient force does not depend on the ultrasound absorption by the particle.
Moreover, it vanishes for a plane traveling wave.

The scattering radiation force reads 
\begin{align}
\nonumber
 {\bm F}_\text{sca} &=  \pi a^2 (ka)^4\biggl[ \frac{4  }{9 }\left(f_0^2 + f_0f_1 + \frac{3f_1^2}{4} \right) 
   \frac{\overline{\bm I}({\bm 0})}{c_0} \\
 &-  \frac{f_1^2 }{6 k}\textrm{Im}[\nabla \cdot \rho_0{\bm v}_\text{i} {\bm v }_\text{i}\mbox{}^*({\bm 0})]
\biggr].
\label{fscat1}
\end{align}
When the incident wave interacts with a rigid (impenetrable) particle, we have $f_0=f_1=1$ and then
\begin{equation}
 {\bm F}_\text{sca}^\textrm{rigid} =  \pi a^2 (ka)^4\biggl[ \frac{11  }{9 } 
   \frac{\overline{\bm I}({\bm 0})}{c_0} 
   -  \frac{1 }{6 k}\textrm{Im}[\nabla \cdot \rho_0{\bm v}_\text{i} {\bm v }_\text{i}^*({\bm 0})]
\biggr].
\label{fscat_rigid}
\end{equation}

Finally, the absorption radiation force is expressed as
\begin{align}
\nonumber
 {\bm F}_\text{abs} &=  \pi a^2 \tilde{\alpha}_\upsilon k a\biggl[\frac{8(1-f_0) }{3}    
    \frac{ \overline{\bm I}({\bm 0})}{c_0} - \frac{12a(1-f_0)}{5( \tilde{\rho}^{-1}_1 + 2)^2} (ka)  \\
  &\times  \textrm{Im}[\nabla \cdot \rho_0{\bm v}_\text{i} {\bm v }_\text{i}\mbox{}^*({\bm 0})]
    \biggr].
    \label{fabs}
%  {\bm F}_\text{abs} =  \frac{8 \pi a^3 \rho_0 \alpha_\upsilon \omega^{\upsilon} }{3 \rho_1c_1}   
%    \overline{\bm I}({\bm 0}).
\end{align}
If the particle is rigid ($f_0=f_1=1$), we have ${\bm F}_\text{abs}^\textrm{rigid} = 0$.
This result is expected since no wave is transmitted into the particle, and thus no absorption occurs.

\section{Acoustic radiation torque}

An acoustic beam can produce a time-averaged torque, known as the radiation torque, on a particle
with respect to its center.
This happens due to the transferring of the  angular momentum of the beam to the particle. 
It has been demonstrated that the radiation torque produced on an absorbing particle by
an arbitrary time-harmonic wave is~\cite{silva:54003}
\begin{equation}
 {\bm N} =  \frac{\pi a^3  I_0}{c_0} \boldsymbol{\tau}
\end{equation}
 where $\boldsymbol{\tau}$ is the
dimensionless radiation torque vector.
Similarly to the radiation force previously analyzed, the dimensionless radiation torque 
vector is given in terms of the beam-shape and the scattering coefficients.
In the Rayleigh scattering limit, the Cartesian component of this vector can be
calculated by keeping the monopole and the dipole scattering coefficients 
in [Eq.~14, ~\cite{silva:54003}].
Accordingly, we have 
\begin{align}
\label{txy}
 \tau_x + i\tau_y &= - \frac{\sqrt{2}}{\pi (ka)^3} \left( \frac{s_1 + s_1^*}{2} + |s_1|^2\right)
 \left(a_1^{-1} a_1^{0*} + a_1^0 a_1^{1*} \right),\\
 \tau_z &=  - \frac{1}{\pi (ka)^3} \left( \frac{s_1 + s_1^*}{2} + |s_1|^2\right)
 \left(\left|a_1^{1}\right|^2 - \left|a_1^{-1}\right|^2 \right).
 \label{tz}
\end{align}
Using the beam-shape coefficients given in (\ref{a22}) into these equations, one finds
the radiation torque as
\begin{equation}
\label{tau_rayleigh}
 {\bm N} = - \pi a^3 \frac{6 i  }{ (k a)^3}  \left( \frac{s_1 + s_1^*}{2} + |s_1|^2\right) \rho_0
  [{\bm v}_\text{i}({\bm 0}) \times {\bm v}_\text{i}^*({\bm 0})].
 \end{equation}
We emphasize that this vector has real components.
 
Substituting the scaled scattering coefficient given in (\ref{s1}) into this equation, we obtain
radiation torque as
\begin{equation}
 {\bm N} = \pi a^3 
 \frac{12 i (1- f_0)\tilde{\alpha}_\upsilon}{5 ( \tilde{\rho}^{-1}_1 + 2)^2}  (ka)^2
%  -\frac{12 i a^2 \rho_0^3\rho_1c_1 \alpha_\upsilon \omega^{\upsilon + 1} }{5p_0^2(\rho_0+2\rho_1)^2 }
\rho_0[ {\bm v}_\text{i}({\bm 0}) \times {\bm v}_\text{i}^*({\bm 0})]. 
 \label{RT}
\end{equation}
This equation shows that no torque is produced in a non-absorbing Rayleigh particle suspended in an inviscid fluid.
Furthermore, the radiation torque is proportional to $(ka)^2$.

\section{Some examples}

In what follows, it is useful to notice that the
the axial averaged intensity is given by
\begin{equation}
\label{Iz}
\overline{I}_z= \frac{1}{2} \textrm{Re}[p v_{\textrm{i},z}\mbox{}^*],
\end{equation}
 while the imaginary part  of the axial component of the momentum flux divergence is expressed as
\begin{equation}
\label{ImDvv}
 \textrm{Im}[\nabla \cdot \rho_0{\bm v}_\textrm{i} {\bm v}_\textrm{i}\mbox{}^*]_z
 =  \textrm{Im}\left[\rho_0\left(\frac{\partial(v_{\text{i},x} v_{\text{i},z}\mbox{}^*)}{\partial x} 
+ \frac{\partial(v_{\text{i},y} v_{\text{i}, z}\mbox{}^*)}{\partial y}\right)\right].
\end{equation}
In turn, these quantities depend on the fluid velocity, which is given, according to Eq.~(\ref{ceq2}),
by 
\begin{equation}
\label{vi_pi}
{\bm v}_\textrm{i} = -\frac{i\nabla p_\textrm{i}}{\rho_0 c_0 k}.  
\end{equation}

\subsection{Plane traveling  wave}

Consider a plane progressive wave along the the $z$-axis.
The pressure amplitude of the plane wave is given by
\begin{equation}
p_\textrm{i}=p_0 e^{i k z}.
\end{equation}
After calculating the axial fluid velocity with Eq.~(\ref{vi_pi}), and 
using this equations into Eqs.~(\ref{Iz}) and (\ref{ImDvv}), we obtain the scattering and the absorption
radiation forces, in Eqs.~(\ref{fscat1}) and (\ref{fabs}), as
\begin{align}
\label{rf_scat_pw}
F_{\text{sca},z} &=  \pi a^2 (ka)^4  \frac{4 I_0}{9 c_0}\left(f_0^2 + f_0f_1 + \frac{3f_1^2}{4} \right), \\
F_{\text{abs},z} &=  \pi a^2 \tilde{\alpha}_\upsilon k a \frac{8I_0(1-f_0) }{3 c_0}.
\label{rf_abs_pw}
\end{align}
The gradient radiation force is zero.
Equation~(\ref{rf_scat_pw}) was previously
obtained by Gorkov.~\cite{gorkov:773}
Furthermore, the result in Eq.~(\ref{rf_abs_pw}) shows the possibility of having the radiation force by a plane
traveling wave exerted on an absorbing particle,
to be  $(ka)^{-3}$ stronger than the force considering a lossless particle.
When $\upsilon=2$, we obtain the ratio $|F_{\text{abs},z}/F_{\text{sca},z}|$ as given in   
Ref.~[Eq.~(52),~~\onlinecite{lofstedt:2027}],
which in turn has an apparent typographical error (an extra factor of $1/3$).

\subsection{Spherically focused beam}

\begin{figure}
 \centering
\includegraphics[scale=.6]{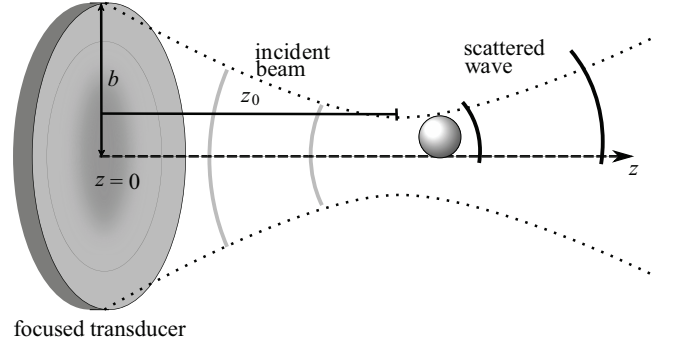}
\caption{Schematic representation of the radiation force exerted by an incident focused beam on a particle
placed at an arbitrary position in a host fluid.
The transducer has aperture $2b$ and focal distance $z_0$.
The axis of the transducer coincides to the $z$ direction.
\label{fig:focusedbeam}
}
\end{figure}

Consider that a spherically focused transducer of diameter $2b$ and curvature radius $z_0$
produced an incident beam to a Rayleigh particle place at the origin of the coordinate system.
The generated ultrasound beam  can be described in the paraxial approximation
if the following conditions are met~\cite{soneson:EL481} $z_0/b\gtrsim 2.5$ and $k b\gtrsim \sqrt{z_0/b}$.
The produced ultrasound beam hits an spherical particle as depicted in Fig.~\ref{fig:focusedbeam}.
In the paraxial approximation, the pressure amplitude is given in cylindrical coordinates
$(\varrho,z)$ by~\cite{lucas:1289}
\begin{align}
\nonumber
 p_\textrm{i}(\varrho,z) &= \frac{i p_0 k e^{i k z}}{z} \exp\left(\frac{i k \varrho^2}{2 z}\right)
 \int_0^b \exp\left[ \frac{i k \varrho'^2}{2}\left(\frac{1}{z} - \frac{1}{z_0}
 \right)
 \right]\\
 &\times J_0\left( \frac{k \varrho \varrho'}{z}\right)\varrho'd\varrho',
 \label{p_focused}
\end{align}
where $J_n$ is the $n$th-order Bessel function.
The pressure given in Eq.~(\ref{p_focused}) may not be valid when $z<0.3 z_0$ as discussed in Ref.~~\onlinecite{lucas:1289}.
In the focal plane and along the axial direction,
the pressure field is reduced, respectively, to
\begin{align}
\label{pr_focused}
 p_\text{i}(\varrho, z_0) &= \frac{i  p_0 b}{\varrho}\exp\left[i k z_0 \left(\frac{\varrho^2}{z_0^2} + 1\right)\right]
 J_1\left(\frac{k \varrho b}{z_0}\right),\\
  p_\text{i}(0,z) &=  \frac{p_0 z_0 e^{i k z}}{z-z_0 } \left[ 1 -
  \exp\left[-\frac{ikb^2}{2}\left( \frac{1}{z_0} - \frac{1}{z}\right)  \right] \right].
\label{pz_focused}
\end{align}

The gradient radiation force may trap a small particle 
in the minimum of the radiation force potential.
Using \textsc{Mathematica} software~\cite{mathematica8},
we calculated the transverse and axial potentials.
However, the obtained expressions are unwieldy and will not be shown here
for the sake of simplicity.
Based on the result, we find that
the radiation force potential on the focal plane has a 
minimum at $\varrho=0$ if
\begin{align}
\nonumber
\frac{\partial^2 U(0,z_0)}{\partial \varrho^2} &= 
- \frac{\pi  p_0^2 k^2 a^3 b^4}{192 \rho_0 c_0^2 z_0^6}\\
&\times [8 k^2 b^2 z_0^2 f_0 +  (3 k^2 b^4  + 192 z_0^2)f_1]>0.
\label{Ur}
\end{align}
Therefore, the condition of transverse trapping in the transducer focus is
\begin{equation}
f_0< - \frac{3(k^2 b^4 + 64 z_0^2)}{8 (k b z_0)^2}f_1.
\label{trap}
\end{equation}
For a denser particle than the host fluid, i.e. $\rho_1>\rho_0$,
we have $f_1>0$.
In this case,  if the particle is also less compressible than the medium, $f_0>0$,
then it cannot be transversely trapped.

The situation is different in the axial direction.
Yet despite the radiation force potential may have a minimum in the focal region, the particle
may not be axially trapped.
This happens because
the absorption and the scattering radiation forces may push the particle
away from the potential minimum.
However, the axial trapping of a particle by a spherical vortex beam
has been theoretically discussed in Ref.~~\onlinecite{baresch2013}.
  
To obtain the  absorption and the scattering radiation forces in the axial direction
 as given in Eqs.~(\ref{fscat1}) and (\ref{fabs}),
 it is necessary to compute the axial components of the averaged intensity and the
momentum flux divergence given, respectively, in Eqs.~(\ref{Iz}) and (\ref{ImDvv}).
Thus, using \textsc{Mathematica} software,~\cite{mathematica8} we find that the axial averaged intensity 
($\varrho=0$) is given by
\begin{equation}
 \overline{I}_z = I_0 \frac{ (4z^2-b^2)z_0^2}{z^2(z-z_0)^2}\sin^2\left[ \frac{kb^2}{4}
\left(\frac{1}{z_0}-\frac{1}{z}\right)\right]
\label{Iz2}
\end{equation}
The expression for  the imaginary part of the  momentum flux divergence was
also calculated. 
Nevertheless, the result is cumbersome and we will not be presented here.

\subsection{Acoustic Bessel beam}

We derive here the radiation force and torque induced by an acoustic Bessel beam  on an absorbing particle
located in the beam axis at ${\bm r}={\bm 0}$.
Assume that the Bessel beam propagates along the $z$-axis.
Hence, the pressure amplitude of this beam is given, in cylindrical coordinates, by
\begin{equation}
\label{bessel}
 p_\text{i} =  p_0  J_n(k \varrho\sin \beta )
 e^{i (n \varphi + k z\cos \beta)},\quad n = 0,\pm 1, \pm 2, \dots,
\end{equation}
where  $\beta$ is the beam's half-cone angle.
The index $n$ is  known as the orbital angular momentum of the beam.
It is worth to notice that the radiation force exerted on a particle due to this beam with $\beta=90^\circ$
has been experimentally performed~\cite{courtney:123508}.

Due to the beam symmetry only the axial radiation force is produced on the particle.
Using Eqs.~(\ref{vi_pi}) and (\ref{bessel}), we find that the axial component of the average incident intensity to the particle 
is given by
$\overline{I}_z({\bm 0}) = I_0 \delta_{n,0} \cos \beta$,
where $\delta_{mn}$ is the Kronecker delta symbol, with $\delta_{nm}=1$ if $n=m$ and $\delta_{nm}=0$, otherwise.
Referring to Eq.~(\ref{ImDvv}) the imaginary part of the momentum flux divergence along the $z$ axis is expressed as
\begin{equation}
 \textrm{Im}[\nabla \cdot \rho_0{\bm v}_\textrm{i} {\bm v}_\textrm{i}^*({\bm 0})]_z=  \frac{I_0}{c_0} b_n k  \cos \beta \sin^2 \beta,
\end{equation}
where $b_0=2$, $b_{\pm 1}=-1$, and $b_n=0$, otherwise.
Therefore, using the averaged intensity and the momentum flux divergence
into Eqs.~(\ref{fscat1}) and (\ref{fabs}), we obtain
\begin{align}
\nonumber
F_{\textrm{abs},z}^{n} &= 
 \pi a^2 \frac{I_0}{c_0}\biggl[\frac{8}{3}(1-f_0) \tilde{\alpha}_\upsilon    
    k a \delta_{n,0} \cos \beta \\
\label{fabs_b0}
    &- \frac{12(1-f_0)b_n}{5( \tilde{\rho}^{-1}_1 + 2)^2}  \tilde{\alpha}_\upsilon(ka)^2 \cos \beta \sin^2 \beta\biggr],\\
\nonumber
 F_{\textrm{sca},z}^{n} &= \frac{\pi a^2 I_0}{c_0}(ka)^4
 \biggl[\frac{4}{9}\left(f_0^2 + f_0f_1 + \frac{3f_1^2}{4} \right) 
    \delta_{n,0} \cos \beta\\
&- \frac{f_1^2}{6}b_n\cos \beta \sin^2 \beta \biggr].
\label{fsca_b0}
\end{align}
The total radiation force exerted on the particle is
$F_{z}^n=F_{\textrm{abs},z}^{n}+F_{\textrm{sca},z}^{n}$.
Note that only  zeroth- and  first-order Bessel beams  produce
axial radiation force on a Rayleigh particle.
Moreover, we verified that Eq.~(\ref{fsca_b0}) for a zeroth-order Bessel beam
is the same as that presented in [Eqs.~(10a) and (15a),~~\cite{marston:3518}].

Now we turn to obtain the acoustic radiation torque by a Bessel vortex beam.
First we calculate the transverse fluid velocity components at the particle center 
using Eqs.~(\ref{vi_pi}) and (\ref{bessel}),
\begin{align}
 v_x({\bm 0}) &= -\frac{i p_0\sin \beta}{2 \rho_0 c_0},\\
 v_y({\bm 0}) &= \frac{p_0\sin \beta}{2 \rho_0 c_0}.
\end{align}
Hence, using this equations into Eq.~(\ref{RT}), we obtain the axial radiation
torque as
\begin{equation}
 N_z = \pi a^3 \frac{I_0}{c_0}
 \frac{6 n \delta_{n,\pm 1} (1- f_0)\tilde{\alpha}_\upsilon } {5 ( \tilde{\rho}^{-1}_1 + 2)^2}  (ka)^2 \sin^2 \beta. 
 \label{bessel_torque}
\end{equation}
Only the first-order Bessel vortex beam can produce on the particle in the Rayleigh scattering limit.
This happens because   radiation torque  is induced by
the dipole mode only as shown in Eqs.~(\ref{txy}) and (\ref{tz}).
The dipole with respect to the beam's axis is not present in all but the first-order Bessel vortex beam.
The term $\sin^2\beta$ present in Eq.~(\ref{bessel_torque}) also appears in the rotational velocity
induced by a first-order Bessel beam on a nonabsorbing particle suspended in a viscous fluid~\cite{marston:045005}.
Note that Eq.~(\ref{bessel_torque}) can also be obtained directly from~[Eq.~14, ~\cite{mitri:026602}].
Moreover, this equation follows immediately from~[Eq.~18, ~\cite{zhang:035601}] 
(with $m = 1$ and $n = 1$ in the associated Legendre function in the notation of that paper), 
combined with the axial radiation torque expression given in~[Eq.~10, ~\cite{zhang:065601}].

\medskip

\section{Numerical results and discussion}

To illustrate how the radiation force and torque are produced,  
we consider two different particles (liquid droplets)
suspended in water ($\rho_0=\unit[1000]{kg/m^3}$, $c_0=\unit[1480]{m/s}$, and $\nu_0=\unit[10^{-6}]{m^2/s}$)
at room temperature.
All incident waves considered here have frequency of $\unit[1]{MHz}$.
The droplets are formed by benzene~\cite{kino:552,crc_handbook:6-177} 
($\rho_1=\unit[870]{kg/m^3}$, $c_1=\unit[1295]{m/s}$,
 $\alpha = \unit[2.21\times 10^{-14}]{Np \: MHz^{-2} \: m^{-1}}$, $\upsilon=2$,
and $\nu_1= \unit[6.94 \times 10^{-7}]{m^2/s}$) 
and olive oil~\cite{coupland:1559} ($\rho_1=\unit[915.8]{kg/m^3}$, $c_1=\unit[1464]{m/s}$, 
$\alpha = \unit[4.10\times 10^{-14}]{Np \: MHz^{-2} \: m^{-1}}$, $\upsilon=2$,
and $\nu_1= \unit[1.00 \times 10^{-4}]{m^2/s}$).
The compressibility and the density contrast factors for the benzene and the
olive oil droplets are $f_0 = -0.5, -0.11$ and $f_1 =-0.09, -0.05$, respectively.
At $\unit[1]{MHz}$ frequency, the inner  viscous boundary layer of the benzene and the olive oil droplets
are, respectively, $\delta_1=\unit[0.47, 5.65]{\mu m}$. 
The outer boundary layer of the droplets is $\delta_0=\unit[0.56]{\mu m}$.
Thus, considering $ka=0.25$ with $a=\unit[58.8]{\mu m}$, 
the inequality in (\ref{ineq_a}) is satisfied.
Hence, we may neglect  the inner and the outer shear wave propagation effects for both 
droplets, since $\delta_0,\delta_1\ll a$.
The droplets were chosen because they are immiscible in water.
However, such small benzene droplets may rapidly dissolve in water 
unless the water bath is previously saturated with benzene.

To produce a focused beam, we consider a spherically focused transducer
with diameter $2b = \unit[50]{mm}$ and curvature radius $z_0=\unit[70]{mm}$.
The transducer operates with intensity $I_0=\unit[33]{W/m^2}$.
We chose these parameters because such transducer could be readily manufactured for
experimental arrangements.
Moreover, the generated ultrasound beam can be described in the paraxial approximation.

\begin{figure}
 \centering
\includegraphics[scale=.5]{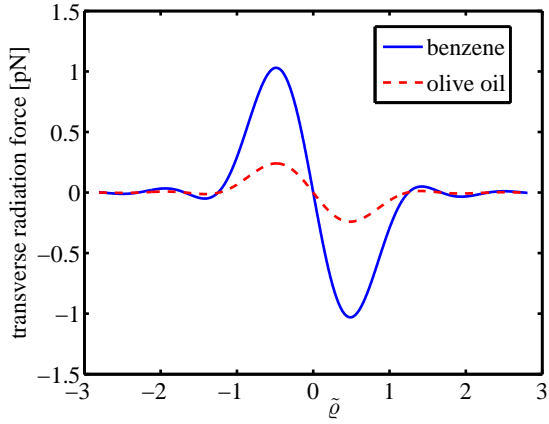}
\caption{
\label{fig:rf_trans}
(Color online) The transverse radiation force generated by the focused transducer  
on the benzene and the olive oil droplets ($ka=0.25$) suspended in water.
The normalized transverse radial distance is $\tilde{\varrho}=\varrho/W_{\unit[-3]{dB} }$.
The transducer parameters are f-number  $z_0/(2b)=1.4$, operation frequency $\unit[1]{MHz}$, focal distance $z_0=\unit[70]{mm}$, 
and the averaged intensity is $I_0=\unit[33]{W/m^2}$.}
\end{figure}

In Fig.~\ref{fig:rf_trans}, we show the transverse radiation force due to the focused transducer
on the benzene and the olive oil droplets.
The force varies with the normalized  transverse radial distance
$\tilde{\varrho}=\varrho/ W_{\unit[-3]{dB} }$,
where~\cite{lucas:1289} $W_{\unit[-3]{dB} }= 1.62 z_0 /(kb) $ is the $\unit[3]{dB}$-width of the focal spot.
Both benzene and olive oil droplets can be  transversely trapped at the transducer focus.

\begin{figure}
 \centering
\includegraphics[scale=.5]{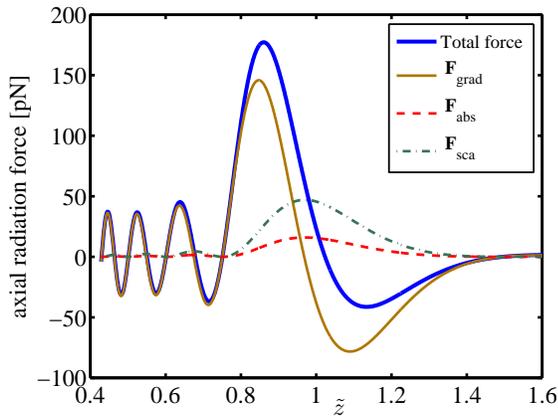}
\caption{\label{fig:rf_axial_benzene}
(Color online) The axial radiation force produced by the focused transducer
on the benzene droplet suspended in water.
The normalized axial coordinate is $\tilde{z}=z/z_0$.
The numerical values of the  parameters used in this evaluation are the same as in Fig.~\ref{fig:rf_trans}.}
\end{figure}

The axial radiation force exerted by the ultrasound focused beam on the benzene droplet
as a function of $\tilde{z}=z/z_0$ is depicted in Fig.~\ref{fig:rf_axial_benzene}.
The droplet could be trapped in the pre-focal zone at  $\tilde{z}=0.46, 0.55, 0.68$,
and at the transducer focal distance $\tilde{z}=1.023$.
These points corresponds to the minima of the radiation force potential $U$.
It should be noticed that the radiation force on the benzene droplet is mostly due to its gradient component.

\begin{figure}
 \centering
\includegraphics[scale=.5]{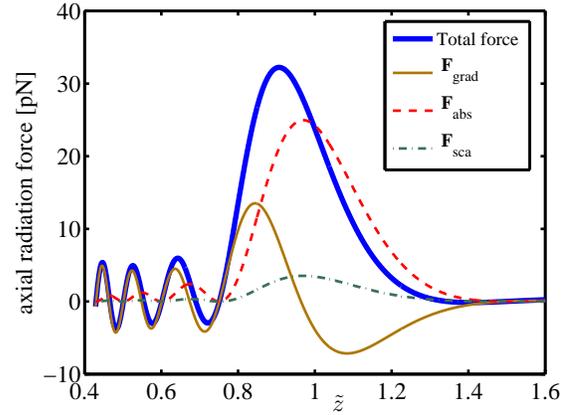}
\caption{\label{fig:rf_axial_silicone}
(Color online) The axial radiation force produced by the focused transducer
on the olive oil droplet suspended in water.
The normalized axial coordinate is $\tilde{z}=z/z_0$.
The evaluation parameters used here are the same as in Fig.~\ref{fig:rf_trans}.}
\end{figure}

In Fig.~\ref{fig:rf_axial_silicone}, we present the axial radiation force exerted on the olive oil droplet
by the focused beam varying with $\tilde{z}=z/z_0$.
The droplet  can be trapped only  in the pre-focal zone at $\tilde{z} = 0.46, 0.55,0.69$.
It can be seen that the absorption radiation force is dominant part in the total radiation force
acting on the olive oil droplet. 
Furthermore, it is worth of noticing that possible entrapment points for both
benzene and olive oil droplets are 
formed in the transducer pre-focal zone ($z<z_0$).
Likewise, the possibility of  particle trapping in the nearfield of a flat transducer was mentioned 
in Ref.~~\onlinecite{mitri:114102}.

%We have numerically verified that the contribution from the momentum flux divergence to the radiation forces by the focused
%beam on both benzene and olive oil droplets in Eqs.~(\ref{fscat1}) and (\ref{fabs}) is smaller than one hundredth of
%the averaged incident intensity contribution.

\begin{figure}
 \centering
\includegraphics[scale=.5]{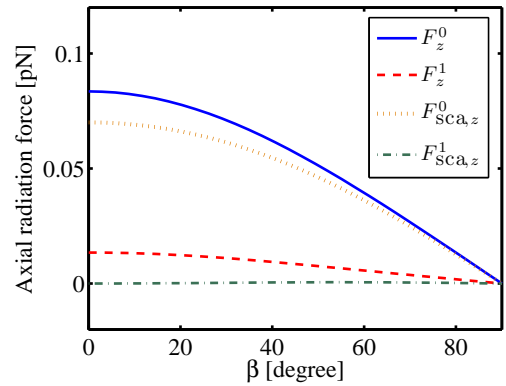}
\caption{\label{fig:rf_bb_axial}
(Color online) The axial radiation force exerted
by a zeroth- and a first-order Bessel vortex beam on the olive oil droplet $(ka=0.25)$ suspended in water.
The radiation force varies with the half-cone angle $\beta$.
The beam averaged intensity is $I_0=\unit[33]{W/m^2}$. }  
\end{figure}

In Fig.~\ref{fig:rf_bb_axial}, the axial radiation force exerted on the olive oil droplet exerted by
a zeroth- and a first-order Bessel vortex beam is shown.
It is clear that the particle absorption plays a major role in the axial radiation force.
The deviation can be as large as $80\%$, but it decreases as the half-cone angle approaches to $90^\circ$.

\begin{figure}
 \centering
\includegraphics[scale=.5]{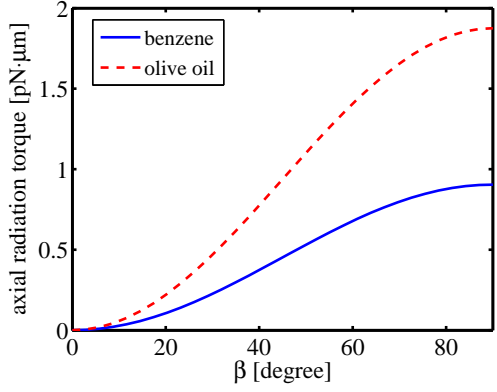}
\caption{\label{fig:rt_axial}
(Color online) The axial radiation torque exerted
by a first-order Bessel vortex beam on the benzene and the olive oil droplets $(ka=0.25)$ suspended in water.
The radiation torque varies with the half-cone angle $\beta$.
The beam averaged intensity is $I_0=\unit[33]{W/m^2}$.}
\end{figure}

Now we present in Fig.~\ref{fig:rt_axial} the axial radiation torque exerted by a first-order Bessel vortex 
beam on the benzene and the olive oil droplets as a function of the half-cone angle $\beta$.
The olive oil droplet  develops a larger radiation torque due to its higher absorption.

\section{Summary and conclusion}

Exact formulas of the radiation force and torque  were provided for any time-harmonic beam 
interacting with an absorbing small particle (in the Rayleigh scattering limit) suspended in an inviscid fluid.
Internal shear viscous effects  were not considered since we assumed that  the inner viscous boundary layer is
much smaller than the particle radius.
Using the developed radiation force theory, the stability of axial and transverse  entrapment of 
a benzene and an olive oil droplet suspended in water by
a spherically focused ultrasound beam were analyzed.
Moreover, the radiation force and torque caused by a zeroth- and a first-order Bessel beam on
the benzene and olive oil droplets in the 
 on-axis configuration  were computed.
 
In conclusion,  the developed closed-form expressions for the radiation force and torque on an absorbing
particle  might be useful in the analysis of trapping stability of single-beam acoustical tweezers.
In a future work, we will take into consideration shear wave propagation effects inside the absorbing 
particle on the acoustic radiation force and torque.

\section*{Acknowledgments}

This work was partially supported by Grant 481284/2012-5 CNPq (Brazilian Agency).

\appendix
\section{Appendix: beam-shape coefficients}

The beam-shape coefficient of a pressure field $p$ is given by~\cite{silva:298}
\begin{equation}
 a_n^m = \int_{\Omega}   \frac{p(k R,\theta,\varphi)}{p_0  j_n(k R)}
Y_n^{m*}(\theta,\varphi)  d\Omega,  n\ge 0, |m|\le n, 
\label{anm}
\end{equation}
where $p_0$ is the peak pressure magnitude, $d\Omega$ is the differential solid angle, 
and $R$ is the radius of a control spherical 
region in which the incident beam propagates.
Expanding the pressure around the origin up to  the second-order approximation,
yields
\begin{equation}
\label{expand_p}
p({\bm r}) = p({\bm 0}) + i \rho_0 c_0 k {\bm r}\cdot {\bm v}({\bm 0})
+ \frac{i\rho_0 c_0 k}{2} {\bm r}\cdot \left[{\bm r} \cdot \nabla {\bm v}({\bm 0})\right],
%\tilde{p}_{i,0} + k R\frac{\partial\tilde{p}_i}{\partial (kR)}\biggl|_{kR=0},
\end{equation}
where  ${\bm v} = v_x{\bm e}_x + v_y{\bm e}_y + v_z {\bm e}_z$ (${\bm e}_i$, $i=x,y,z$
 are the Cartesian unit-vectors).
Substituting Eq.~(\ref{expand_p}) into Eq.~(\ref{anm}) along with 
${\bm r} = R (\sin \theta \cos\varphi {\bm e}_x + 
\sin \theta \sin\varphi {\bm e}_y + \cos\theta{\bm e}_z )$ and $R\rightarrow 0$, 
we obtain the beam-shape coefficients  up to the quadrupole approximation as
%Equations verified with Mathematica (file bsc.nb)
\begin{align}
\nonumber
 a_0^0 &= \frac{\sqrt{4\pi}}{p_0}  p({\bm 0}),\\
\nonumber
 a_1^{\mp 1} &= \frac{\rho_0 c_0}{p_0} \sqrt{6 \pi} [\pm i v_{x}({\bm 0}) + v_{y}({\bm 0})],\\
 \nonumber
 a_1^0 &= \frac{ 2 i \rho_0 c_0}{p_0}  \sqrt{3 \pi} v_{z}({\bm 0}),\\
 %\nonumber
 %a_1^1 &= \frac{\rho_0 c_0}{p_0} \sqrt{6 \pi} [-i v_{x}(0) + v_{y}(0)],\\
  \label{a22}
 a_2^{\mp 2} &= \frac{\rho_0 c_0}{k p_0}\sqrt{\frac{15\pi}{2}} \left[ i\left(
 \frac{\partial v_{x}({\bm 0})}{\partial x} - \frac{\partial v_{y}({\bm 0})}{\partial y}\right) \pm  2 \frac{\partial v_{x}({\bm 0})}{\partial y} \right] ,\\
\nonumber
 a_2^{\mp 1} &= \frac{\rho_0 c_0}{k p_0} \sqrt{30\pi}  \left[
 \pm i \frac{\partial v_{x}({\bm 0})}{\partial z}
   + \frac{\partial v_{y}({\bm 0})}{\partial z}
\right],\\
 \nonumber
 a_2^0 &= -\frac{ i \rho_0 c_0 }{k p_0} \sqrt{5 \pi} \left[ 
   \frac{\partial v_{x}({\bm 0})}{\partial x} + \frac{\partial v_{y}({\bm 0})}{\partial y} - 2 \frac{\partial v_{z}({\bm 0})}{\partial z} \right].\\
 \nonumber
\end{align}  

%\bibliographystyle{jasasty.bst}
%\bibliography{RF_references}

\end{document}